\documentclass[final,3p,twocolumn,times]{elsarticle}
\usepackage[utf8]{inputenc}
\usepackage[T1]{fontenc}
\usepackage{amsmath}
\usepackage{amssymb,amsfonts,amsthm}
\usepackage[all,cmtip]{xy}
\journal{}

\theoremstyle{definition}

\theoremstyle{remark}

\catcode`,\active

\catcode`\,12

\catcode`,\active

\catcode`\,12

\newcommand{\pd}{\partial}

\begin{document}

\begin{frontmatter}



\title{The Dunkl-Coulomb problem in the plane}


\author[UdeM]{Vincent X. Genest}
\ead{genestvi@crm.umontreal.ca}
\author[UdeM]{Andr\'eanne Lapointe}
\ead{andreanne.lapointe.2@umontreal.ca}
\author[UdeM]{Luc Vinet}
\ead{vinetl@crm.umontreal.ca}

\address[UdeM]{Centre de Recherches Math\'ematiques et D\'epartement de Physique, Universit\'e de Montr\'eal, CP 6128, Succursale Centre-Ville, Montr\'eal (QC), H3C 3J7, Canada}
\begin{abstract}
The Dunkl-Coulomb system in the plane is considered. The model is defined in terms of the Dunkl Laplacian, which involves reflection operators, with a $r^{-1}$ potential. The system is shown to be maximally superintegrable and exactly solvable. The spectrum of the Hamiltonian is derived algebraically using a realization of $\mathfrak{so}(2,1)$ in terms of Dunkl operators. The symmetry operators generalizing the Runge-Lenz vector are constructed. On eigenspaces of fixed energy, the invariance algebra they generate is seen to correspond to a deformation of $\mathfrak{su}(2)$ by reflections. The exact solutions are given as products of Laguerre polynomials and Dunkl harmonics on the circle.
\end{abstract}
\end{frontmatter}
\section{Introduction}
\noindent
The purpose of this letter is to study the Dunkl-Coulomb system in the plane governed by the Hamiltonian
\begin{align}
\label{Hamiltonian}
\mathcal{H}=-\frac{1}{2}\nabla_{D}^2+\frac{\alpha}{r},
\end{align}
with $r^2=x_1^2+x_2^2$. In the above $\nabla_{D}^2$ stands for the two-dimensional Dunkl-Laplace operator
\begin{align*}
\nabla_{D}^2=\mathcal{D}_1^2+\mathcal{D}_2^2,
\end{align*}
defined in terms of the Dunkl derivative \cite{Dunkl-1989-01}
\begin{align}
\label{Dunkl-D}
\mathcal{D}_i=\pd_{x_i}+\frac{\mu_i}{x_i}(1-R_i),\quad i=1,2,
\end{align}
where $R_i$ is the reflection operator
\begin{align*}
R_i f(x_i)=f(-x_i),
\end{align*}
with respect to the $x_i=0$ axis and where $\mu_i>0$ are real parameters. It will be shown that the model described by \eqref{Hamiltonian} is both maximally superintegrable and exactly solvable. The spectrum of the Hamiltonian will be derived algebraically using a realization of $\mathfrak{so}(2,1)$ with Dunkl operators. The constants of motion, which are analogous to the Runge-Lenz vector, will be constructed and it shall be seen that these symmetries close rationally to a deformation of $\mathfrak{su}(2)$ by reflections. The exact solutions of the model will be given in terms of Laguerre polynomials and the Dunkl harmonics on the circle.

Generically, a quantum system with $d$ degrees of  freedom determined by a Hamiltonian $H$ is said to be maximally superintegrable if it admits $2d-1$ algebraically independent symmetry operators $S_j$ that commute with the Hamiltonian
\begin{align*}
[H,S_j]=0,\quad j=1,\ldots,2d-1,
\end{align*}
where one of the symmetries is the Hamiltonian itself. In this case, it is impossible for all the symmetries to be in involution and thus they generate non-Abelian invariance (or symmetry) algebras. Classical examples of superintegrable systems include the two-dimensional harmonic oscillator and the Coulomb-Kepler system.

In view of their numerous applications (see \cite{Miller-2013-10} for a recent review) there is a continued interest in enlarging the set of documented superintegrable systems. Progress in this regard was made recently with the introduction of a  series of novel superintegrable systems involving reflections \cite{Genest-2014-1, Genest-2013-12-1, Genest-2013-04, Genest-2013-09, Genest-2013-07}. These models are described by Hamiltonians involving the Dunkl-Laplace operator. They generalize the standard and singular oscillators in the plane and on the 2-sphere. Interestingly, these models have also been shown to serve as showcases for the recently introduced  $-1$ orthogonal polynomials \cite{Genest-2013-09-02,Genest-2013-02-1,Tsujimoto-2012-03,Tsujimoto-2013-03-01,Vinet-2012-05}.

The investigation and characterization of superintegrable systems with reflections is pursued here with the study of the Dunkl-Coulomb problem. As shall be seen, this model exhibits many interesting properties.
\section{Algebraic derivation of the spectrum}
\noindent
The spectrum of the Hamiltonian \eqref{Hamiltonian} governing the Dunkl-Coulomb system can be obtained algebraically using its underlying $\mathfrak{so}(2,1)$ dynamical symmetry. We first discuss the relevant realization of $\mathfrak{so}(2,1)$ introduced in \cite{Kobayashi-2012} (see also \cite{DeBie-2011}) and then establish the connection between this realization and the Dunkl-Coulomb problem.
\subsection{A realization of $\mathfrak{so}(2,1)$ with Dunkl operators}
\noindent
Let $E$ denote the dilation operator
\begin{align*}
E=x_1 \pd_{x_1}+x_2 \pd_{x_2}+(\mu_1+\mu_2+1/2),
\end{align*}
and define
\begin{align}
\label{Realization}
\begin{aligned}
L_0&=-r\,\nabla^2_{D}+\frac{r}{4},
\\
L_{\pm}&=-r\,\nabla^2_{D}-\frac{r}{4}\pm E.
\end{aligned}
\end{align}
It is directly checked that the operators $L_0$, $L_{\pm}$ satisfy the defining relations of the $\mathfrak{so}(2,1)$ algebra
\begin{align*}
[L_0, L_{\pm}]=\pm L_{\pm},\quad [L_{+},L_{-}]=-2L_{0}.
\end{align*}
In the realization \eqref{Realization}, the $\mathfrak{so}(2,1)$ Casimir operator 
\begin{align*}
C=L_0^2-\frac{1}{2}(L_{+}L_{-}+L_{-}L_{+}),
\end{align*}
is found to have the expression
\begin{align}
\label{Casimir}
C=J_3^2+(\mu_1 R_1+\mu_2 R_2)^2-1/4,
\end{align}
where $J_3$ is the Dunkl angular momentum generator
\begin{align}
\label{Dunkl-Angular}
J_3=-i(x_1 \mathcal{D}_2- x_2 \mathcal{D}_1).
\end{align}
In irreducible representations of the positive-discrete series of $\mathfrak{so}(2,1)$, the Casimir operator acts as a multiple of the identity
\begin{align}
\label{Form}
C=\nu(\nu-1),\quad \nu>0,
\end{align}
and the eigenvalues of the generator $L_0$, denoted by $\lambda_{L_0}(\ell)$, are of the form
\begin{align}
\label{Eigen-Zero}
\lambda_{L_0}(\ell)=\ell+\nu,
\end{align}
where $\ell$ is a non-negative integer (see \cite{Vilenkin-1991}). 

In view of  \eqref{Casimir} and considering representations of the positive-discrete series, it follows from \eqref{Eigen-Zero} that the spectrum of the operator $L_0$ in the realization \eqref{Realization} can be obtained if the eigenvalues of the operator \eqref{Dunkl-Angular}, which give the decomposition of the representation \eqref{Realization} in irreducible components, are known. In \cite{Genest-2013-04} and \cite{Genest-2013-09} the spectrum of \eqref{Dunkl-Angular} has been found. It splits into two sectors corresponding to the possible eigenvalues of the operator $R_1R_2$, which commutes with \eqref{Dunkl-Angular}. In the sector $R_1R_2=+1$, the eigenvalues of  \eqref{Dunkl-Angular}, denoted by $\lambda_{J_3}^{+}(m)$, read
\begin{align}
\label{Eigen-1}
\lambda_{J_3}^{+}(m)=\pm 2\sqrt{m(m+\mu_1+\mu_2)},
\end{align}
where $m$ is a non-negative integer. In the sector $R_1R_2=-1$, the eigenvalues $\lambda_{J_3}^{-}(k)$ of \eqref{Dunkl-Angular} are given by
\begin{align}
\label{Eigen-2}
\lambda_{J_3}^{-}(k)=\pm 2\sqrt{(k+\mu_1)(k+\mu_2)},
\end{align}
where $k$ is a positive half-integer, i.e. $k\in \{1/2, 3/2,\ldots\}$. Note that the eigenvalue $\lambda_{J_3}^{+}(0)=0$ is non-degenerate. Upon combining \eqref{Eigen-1} and \eqref{Eigen-2}, one finds that the eigenvalues of the Casimir operator \eqref{Casimir} in the realization \eqref{Realization} are of the form \eqref{Form} with
\begin{align*}
\nu(n)=2n+\mu_1+\mu_2+1/2,
\end{align*}
where  $n$ is a non-negative integer for $R_1R_2=+1$ and a positive half-integer for $R_1R_2=-1$. The  irreducible $\mathfrak{so}(2,1)$ representations contained in the realization \eqref{Realization} thus belong to the positive-discrete series and one has the expression
\begin{align}
\label{Spectrum-1}
\lambda_{L_0}(\ell,n)=\ell+2n+\mu_1+\mu_2+1/2,
\end{align}
for the spectrum of the compact generator $L_0$.
\subsection{Connection with the Dunkl-Coulomb problem}
\noindent
The connection between the realization \eqref{Realization} and the Dunkl-Coulomb system works in a similar fashion to the standard case without reflections (see for example \cite{Barut-1988, Vinet-1985, Floreanini-1993-09}). It is established as follows. Consider the Schr\"odinger equation for the Hamiltonian  \eqref{Hamiltonian} of the Dunkl-Coulomb system
\begin{align*}
\left(-\frac{1}{2}\nabla^2_{D}+\frac{\alpha}{r}\right)\Psi_{\mathcal{E}}=\mathcal{E}\,\Psi_{\mathcal{E}}.
\end{align*}
For bound states ($\mathcal{E}<0$), one can multiply both sides of the above equation by $r$ to find
\begin{align*}
\left(-\frac{r}{2}\nabla^2_{D}-\mathcal{E}\,r\right)\Psi_{\mathcal{E}}=-\alpha\;\Psi_{\mathcal{E}}.
\end{align*}
Upon rescaling  the coordinates according to
\begin{align*}
x_i\rightarrow x_i/\sqrt{-8\,\mathcal{E}},
\end{align*} 
the eigenvalue equation is transformed into
\begin{align*}
L_0 \Psi_{\mathcal{E}}=-\frac{2\alpha}{\sqrt{-8\,\mathcal{E}}}\,\Psi_{\mathcal{E}}.
\end{align*}
From the formula \eqref{Spectrum-1} giving the eigenvalues of $L_0$ in the realization \eqref{Realization}, the energy spectrum of the Hamiltonian \eqref{Hamiltonian} for the Dunkl-Coulomb system is thereby immediately found to have the expression
\begin{align}
\label{Energy-Spectrum}
\mathcal{E}(\ell,n)=\frac{-\alpha^2}{2(\ell+2n+\mu_1+\mu_2+1/2)^2},
\end{align}
where $\ell$ is a non-negative integer and where $n$ is a non-negative integer in the sector $R_1R_2=+1$ and a positive half-integer in the sector $R_1R_2=-1$. Note that the Hamiltonian \eqref{Hamiltonian} commutes with both reflection operators, thus confirming that $\mathcal{H}$ and $R_1R_2$ can be diagonalized simultaneously.

As is seen from \eqref{Dunkl-D}, when $\mu_1=\mu_2=0$ the Dunkl-Coulomb system reduces to the standard Coulomb-Kepler problem. It is easily checked that in this case one recovers from \eqref{Energy-Spectrum} the standard expression for the energy spectrum of the Coulomb system.
\section{Superintegrability and invariance algebra}
\noindent
To establish the superintegrability of the Dunkl-Coulomb system in the plane, one needs to find two algebraically independent operators that commute with the Hamiltonian. This can be done in the following way. Let $\mathcal{A}_1$, $\mathcal{A}_2$ be the operators defined as
\begin{align}
\label{Symmetries}
\begin{aligned}
\mathcal{A}_1&=\frac{x_1}{r}-\frac{\mu_1}{\alpha}\,\mathcal{D}_1\,R_1-\frac{1}{2\alpha}\{\mathcal{J},\mathcal{D}_2\},
\\
\mathcal{A}_2&=\frac{x_2}{r}-\frac{\mu_2}{\alpha}\,\mathcal{D}_2\,R_2+\frac{1}{2\alpha}\{\mathcal{J},\mathcal{D}_1\},
\end{aligned}
\end{align}
where 
\begin{equation}
\mathcal{J}=(x_1\mathcal{D}_2-x_2 \mathcal{D}_1)=i J_3,
\end{equation}
and where $\{x,y\}=xy+yx$ stands for the anticommutator. A direct computation shows that $\mathcal{A}_1$, $\mathcal{A}_2$ and $\mathcal{J}$ are constants of motion for the Dunkl-Coulomb system, i.e.
\begin{align}
[\mathcal{H},\mathcal{J}]=[\mathcal{H},\mathcal{A}_1]=[\mathcal{H},\mathcal{A}_2]=0,
\end{align}
with $\mathcal{H}$ given by \eqref{Hamiltonian}. Since \eqref{Hamiltonian} commutes with the reflections $R_1$, $R_2$, these operators are also (discrete) symmetries of the Dunkl-Coulomb Hamiltonian. The constants of motion $\mathcal{A}_1$ and $\mathcal{A}_2$ are analogous to the components of the Runge-Lenz vector for the standard Coulomb-Kepler system in two dimensions. 

The invariance algebra generated by the constants of motion $\mathcal{A}_1$, $\mathcal{A}_2$, $\mathcal{J}$ is found to be
\begin{align}
\label{Symm-Algebra}
\begin{aligned}
{}
[\mathcal{A}_1,\mathcal{A}_2]&=-\frac{2}{\alpha^2}\mathcal{H}\mathcal{J},
\\
[\mathcal{A}_1,\mathcal{J}]&=\mathcal{A}_2(1+2\mu_1 R_1),
\\
[\mathcal{J},\mathcal{A}_2]&=\mathcal{A}_1(1+2\mu_2 R_2),
\end{aligned}
\end{align}
and the commutation relations involving the reflections are seen to take the form
\begin{align}
\begin{aligned}
&\{\mathcal{J},R_1\}=0,\quad &&\{\mathcal{J},R_2\}=0,
\\
&\{\mathcal{A}_1,R_1\}=0,\quad &&[\mathcal{A}_1,R_2]=0,
\\
&\{\mathcal{A}_2,R_2\}=0,\quad &&[\mathcal{A}_2,R_1]=0,
\end{aligned}
\end{align}
with $[R_1,R_2]=0$. The symmetry algebra \eqref{Symm-Algebra} has the Casimir operator
\begin{multline}
\label{Cas-2}
Q=\mathcal{A}_1^2+\mathcal{A}_2^2+\frac{2\mathcal{H}}{\alpha^2}\mathcal{J}^2
\\
-\frac{2\mathcal{H}}{\alpha^2}(\mu_1R_1+\mu_2 R_2+2\mu_1\mu_2 R_1R_2),
\end{multline}
which commutes with all symmetries. A direct computation shows that the Casimir operator $Q$ is related to the Hamiltonian by
\begin{align}
Q=\frac{\mathcal{H}}{\alpha^2}(2\mu_1^2+2\mu_2^2+1/2)+1.
\end{align}
On an eigenspace of $\mathcal{H}$ with a given value of the energy $\mathcal{E}$, one can introduce the renormalized operators
\begin{align*}
J_1=\sqrt{\frac{\alpha^2}{-2\mathcal{H}} }\,\mathcal{A}_1, \quad J_2=\sqrt{\frac{\alpha^2}{-2\mathcal{H}}}\,\mathcal{A}_2.
\end{align*}
In terms of the operators $J_1$, $J_2$ and $J_3$ (with $J_3$ given by \eqref{Dunkl-Angular}), the invariance algebra of the Dunkl-Coulomb systems becomes
\begin{align}
\label{Sym-3}
\begin{aligned}
{}
[J_1,J_2]&=i J_3,
\\
[J_2,J_3]&=i J_1(1+2\mu_2 R_2),
\\
[J_3,J_1]&=i J_2(1+2\mu_1 R_1),
\end{aligned}
\end{align}
with
\begin{gather*}
\{J_1,R_1\}=\{J_2,R_2\}= \{J_3,R_1\}=\{J_3,R_2\}=0,
\\
[J_1,R_2]=[J_2,R_1]=[R_1,R_2]=0.
\end{gather*}
The Casimir operator can then be expressed as
\begin{align*}
Q=J_1^2+J_2^2+J_3^2+\mu_1 R_1+\mu_2 R_2+2\mu_1\mu_2 R_1R_2.
\end{align*}
It is apparent from \eqref{Sym-3} that the symmetry algebra of the Dunkl-Coulomb problem corresponds to deformation of $\mathfrak{su}(2)$ by reflections.  The standard $\mathfrak{su}(2)$ commutation relations are recovered when one takes $\mu_1=\mu_2=0$.  The physical representations of the algebra \eqref{Sym-3} can be obtained in a straightforward fashion. The computations are somewhat tedious and the action of the generators on the basis states are involved. The corresponding formulas are not reproduced here but can be found in \cite{Lapointe-2014}.
\section{Exact solutions}
\noindent
The Schr\"odinger equation associated to the Dunkl-Coulomb Hamiltonian
\begin{align}
\label{Sch-1}
\mathcal{H}\,\Psi=\mathcal{E}\,\Psi,
\end{align}
can be exactly solved using separation of variables in polar coordinates 
$$x_1=r \cos \phi \quad x_2=r\sin \phi.
$$
Since the Hamiltonian commutes with both reflections, it is convenient to look for the joint eigenfunctions of  $\mathcal{H}$, $R_1$ and $R_2$. In polar coordinates, the Hamiltonian \eqref{Hamiltonian} has the expression
\begin{align*}
\mathcal{H}=A_{r}+\frac{1}{r^2} B_{\phi},
\end{align*}
where $A_{r}$ is given by
\begin{align*}
A_r=-\frac{1}{2}\pd_{r}^2-\frac{1}{2r}(1+2\mu_1+2\mu_2)\,\pd_{r}+\frac{\alpha}{r},
\end{align*}
and where $B_{\phi}$ reads
\begin{multline*}
B_{\phi}=-\frac{1}{2}\pd_{\phi}^2+(\mu_1\,\mathrm{tg}\,\phi-\mu_2\,\mathrm{ctg}\,\phi)\pd_{\phi}
\\
+\frac{\mu_1(1-R_1)}{2\cos^2\phi}+\frac{\mu_2(1-R_2)}{2\sin^2\phi}.
\end{multline*}
Upon taking $\Psi=R(r)\,\Phi(\phi)$, one finds that \eqref{Sch-1}  becomes
\begin{subequations}
\begin{align}
\label{A}
\left(A_r-\mathcal{E}+\frac{m^2}{2\rho^2}\right)R(r)&=0,
\\
\label{B}
\left(B_{\phi}-\frac{m^2}{2}\right)\Phi(\phi)&=0,
\end{align}
\end{subequations}
where $m^2/2$ is the separation constant. 

The equation \eqref{B} is similar to the one arising in the study of the two-dimensional Dunkl harmonic oscillator system \cite{Genest-2013-04}. The solutions are labeled by the quantum numbers $(e_1,e_2)$ corresponding to the eigenvalues $(1-2e_1,1-2e_2)$ of the reflection operators $(R_1,R_2)$ ($e_i\in\{0,1\}$). They are given by
\begin{multline}
\label{Polar}
\Phi_{n}^{(e_1,e_2)}(\phi)=\eta_{n}^{(e_1,e_2)}
\\
\times
\cos^{e_1}\phi\,\sin^{e_2}\phi\;P_{n-e_1/2-e_2/2}^{(\mu_1-1/2+e_1,\,\mu_2-1/2+e_2)}(-\cos 2\phi),
\end{multline}
where $P_{n}^{(\alpha,\beta)}(x)$ are the Jacobi polynomials \cite{Koekoek-2010}. When $(e_1,e_2)\in \{(0,0), (1,1)\}$, $n$ is a non-negative integer and when $(e_1,e_2)\in \{(1,0), (0,1)\}$, $n$ is a positive half-integer. The normalization constant
\begin{multline*}
\eta_{n}^{(e_1,e_2)}=\sqrt{\left(\frac{2n+\mu_1+\mu_2}{2}\right)\,\left(n-\frac{e_1+e_2}{2}\right)!}
\\
\times \sqrt{\frac{\Gamma(n+\mu_1+\mu_2+\frac{e_1+e_2}{2})}{\Gamma(n+\mu_1+\frac{1+e_1-e_2}{2})\;\Gamma(n+\mu_2+\frac{1+e_2-e_1}{2})}},
\end{multline*}
where $\Gamma(x)$ is the classical gamma function \cite{Arfken-2012}, ensures that the wavefunctions satisfy the orthogonality relation
\begin{multline*}
\int_{0}^{2\pi} \Phi_{n}^{(e_1.e_2)}(\phi)\,\Phi_{n'}^{(e_1',e_2')}(\phi)\;\rvert \cos \phi \,\rvert^{2\mu_1}  \rvert \sin \phi \,\rvert^{2\mu_2}\,\mathrm{d}\phi
\\
=\delta_{nn'}\delta_{e_1e_1'}\delta_{e_2e_2'}.
\end{multline*}
Note that $B_{\phi}$ is related to $J_3$ as follows:
\begin{align*}
J_3^2=2B_{\phi}+2\mu_1\mu_2(1-R_1R_2).
\end{align*}
Furthermore, it is readily seen that the wavefunctions \eqref{Polar} correspond to the so-called Dunkl harmonics on the circle (see \cite{Dunkl-2001}).

For the solutions \eqref{Polar}, the separation constant has the expression 
$$m^2=4n(n+\mu_1+\mu_2).
$$
Upon substituting this value in the equation \eqref{A}, the radial wavefunctions can be obtained. One finds
\begin{multline}
\label{Radial}
R_{\ell,n}(r)=\xi_{\ell,n}
\\
\times e^{-\beta \,r/2}(\beta\,r)^{2n} \;L_{\ell}^{(4n+2\mu_1+2\mu_2)}(\beta\,r)
\end{multline}
where $L_{n}^{(\alpha)}(x)$ are the Laguerre polynomials \cite{Koekoek-2010} and where $\beta$ is given by
\begin{align*}
\beta=\sqrt{-8\;\mathcal{E}(\ell,n)},
\end{align*}
with $\mathcal{E}(\ell,n)$ given by \eqref{Energy-Spectrum}. The normalization factor
\begin{multline*}
\xi_{\ell,n}=\sqrt{\frac{\ell !}{\Gamma(\ell+4n+2\mu_1+2\mu_2+1)}}
\\
\times \sqrt{\frac{\beta^{2\mu_1+2\mu_2+2}}{(2\ell+4n+2\mu_1+2\mu_2+1)}},
\end{multline*}
ensures that one has
\begin{align*}
\int_{0}^{\infty} R_{\ell,n}(r)\;R_{\ell',n}(r)\;r^{2\mu_1+2\mu_2+1}\mathrm{d}r=\delta_{\ell\ell'}.
\end{align*}

In view of the above results, the eigenfunctions $\Psi_{\ell,n}(r,\phi)$ of the Dunkl-Coulomb Hamiltonian \eqref{Hamiltonian} corresponding to the energy values $\mathcal{E}(\ell,n)$ given by \eqref{Energy-Spectrum} have the expressions
\begin{align}
\label{Full-Wave}
\Psi_{\ell,n}(r,\phi)=R_{n,\ell}(r)\;\Phi_{n}^{(e_1,e_2)}(\phi),
\end{align}
where the radial and angular parts are respectively given  by \eqref{Polar} and \eqref{Radial}. The wavefunctions  are orthogonal under the scalar product
\begin{multline*}
\langle f, g \rangle =
\\
\int_{0}^{\infty}\int_{0}^{2\pi} f^{*}(r,\phi)\,g(r,\phi)\;|r\cos\phi|^{2\mu_1}|r\sin\phi|^{2\mu_2} \,r\,\mathrm{d}r\,\mathrm{d}\phi,
\end{multline*}
with respect to which the Dunkl-Laplace operator is self-adjoint \cite{Genest-2013-04}. It is easily checked that if one takes $\mu_1=\mu_2=0$, the usual wavefunctions of the 2D Coulomb-Kepler system are recovered from \eqref{Full-Wave}.

The Schr\"odinger equation associated to the Dunkl-Coulomb system does not seem to admit separation of variables in any other coordinate system.This is in contradistinction with the classical two-dimensional Coulomb-Kepler Hamiltonian whose separability in parabolic coordinates can be traced back to the existence of the Runge-Lenz vector (see for example \cite{Kalnins-1996}). Let us remark also that the Dunkl-Coulomb problem is not related to the Dunkl oscillator by the Levi-Civita transformation \cite{Boiteux-1982, Levi-Civita-1920} or the coupling constant metamorphosis \cite{Kalnins-Post-Miller-2010}.

\section{Conclusion}
\noindent
In this letter we studied the Dunkl-Coulomb system in the plane. We showed that this model is both superintegrable and exactly solvable. The constants of motion were obtained and the symmetry algebra they satisfy was given. In addition, the separated solutions were given explicitly in polar coordinates.

This adds to the expanding set of superintegrable systems in two dimensions with reflections. A classification of those systems would now deserve attention. In the scalar case and without reflections, it is known that all superintegrable systems whose constants of motion are of second degree in the momenta can be obtained by limits and contractions of the generic model on the 2-sphere \cite{Genest-2013-tmp-1,Kalnins-2013-05}. It would be enlightening to show that similarly the Dunkl-Coulomb problem is a limit of a more generic system.
\section*{Acknowledgements}
\noindent
V.X.G. holds an Alexander-Graham-Bell fellowship from the Natural Sciences and Engineering Research Council of Canada (NSERC). The research of L.V. is supported in part by NSERC. 

\end{document}